\documentclass[conference]{IEEEtran}

\usepackage{cite}
\usepackage{amsmath,amssymb,amsfonts}
\usepackage{algorithmic}
\usepackage{graphicx}
\usepackage{textcomp}
\usepackage{tikz}

\usepackage[hidelinks]{hyperref}
\usepackage{url}
\usepackage{booktabs}
\usepackage{cleveref}
\usepackage[acronyms]{glossaries}
\usepackage{subfig}
\usepackage{svg}
\usepackage{enumitem}
\usepackage{stfloats}
\usepackage[linesnumbered,ruled]{algorithm2e}

\usepackage{soul}
\usepackage[table,xcdraw,dvipsnames]{xcolor}

\usepackage{multirow}
\usepackage{fontawesome5}
\usepackage{accsupp}
\usepackage{tabulary}
\usepackage{calc}

\newcommand{\secure}{%
    \BeginAccSupp{ActualText=secure connection}%
    \textcolor{ForestGreen}{\faCheckCircle}%
    \EndAccSupp{}%
}
\newcommand{\shield}{%
     \BeginAccSupp{ActualText=blocked connection}%
     \textcolor{NavyBlue}{\faShield*}%
     \EndAccSupp{}%
}
\newcommand{\http}{%
    \BeginAccSupp{ActualText=http connection}%
    \textcolor{BurntOrange}{\faExclamationCircle}%
    \EndAccSupp{}%
}
\newcommand{\untrusted}{%
    \BeginAccSupp{ActualText=untrusted connection}%
    \textcolor{BurntOrange}{\faExclamationTriangle}%
    \EndAccSupp{}%
}
\usepackage{pifont}
\newcommand{\yes}{%
    \BeginAccSupp{ActualText=fulfilled}%
    \Large\bfseries\textcolor{ForestGreen}{\ding{51}}%
    \EndAccSupp{}%
}
\newcommand{\no}{%
    \BeginAccSupp{ActualText=unsatisfied}%
    \Large\bfseries\textcolor{Red}{\ding{55}}%
    \EndAccSupp{}%
}

\newcommand{\circled}[2][]{%
    \tikz[baseline=(char.base)]{
        \node[
            circle,
            draw,
            inner sep=1.4pt,
            fill=#1
        ] (char) {#2};
    }%
}
\definecolor{lightgrey}{HTML}{dddddd}
\newcommand{\circledintext}[2][]{%
    \tikz[baseline={([yshift=-0.2ex]char.base)}]{
        \node[
            circle,
            draw,
            inner sep=.8pt,
            fill=#1
        ] (char) {\scriptsize #2};
    }%
}

\usepackage{caption}
\crefname{figure}{Fig.}{Figs.}
\Crefname{figure}{Fig.}{Figs.}

\setacronymstyle{short-long}
\newacronym[longplural={Certificate Authorities}]{ca}{CA}{Certificate Authority}
\newacronym{caa}{CAA}{Certification Authority Authorization}
\newacronym{csp}{CSP}{Content Security Policy}
\newacronym{ct}{CT}{Certificate Transparency}
\newacronym{dane}{DANE}{DNS-based Authentication of Named Entities}
\newacronym{dns}{DNS}{Domain Name System}
\newacronym{dnssec}{DNSSEC}{DNS Security Extension}
\newacronym{doh}{DoH}{DNS-over-HTTPS}
\newacronym{tld}{TLD}{top-level domain}
\newacronym{ip}{IP}{Internet Protocol}
\newacronym{http}{HTTP}{Hypertext Transfer Protocol}
\newacronym{https}{HTTPS}{Hypertext Transfer Protocol Secure}
\newacronym{hsts}{HSTS}{HTTP Strict Transport Security}
\newacronym{pkp}{PKP}{Public Key Pinning}
\newacronym{ssl}{SSL}{Secure Sockets Layer}
\newacronym{tls}{TLS}{Transport Layer Security}
\newacronym{ntp}{NTP}{Network Time Protocol}
\newacronym{rtt}{RTT}{Round-Trip Time}
\newacronym{xss}{XSS}{Cross-Site Scripting}
\newacronym{zltp}{ZLTP}{Zero-Leakage Transfer Protocol}
\newacronym{mta-sts}{MTA-STS}{Mail Transfer Agent Strict Transport Security}
\newacronym{sshfp}{SSHFP}{Secure Shell Fingerprint}
\newacronym{vpn}{VPN}{Virtual Private Network}
\newacronym{mdns}{mDNS}{Multicast DNS}

\title{Securing the Web with HSTS-Enforced}

\author{
\IEEEauthorblockN{Aaron van Diepen}
\IEEEauthorblockA{
\textit{Delft University of Technology}\\
Delft, The Netherlands}
\and
\IEEEauthorblockN{Adrian Zapletal}
\IEEEauthorblockA{
\textit{Delft University of Technology}\\
Delft, The Netherlands}
\and
\IEEEauthorblockN{Fernando Kuipers}
\IEEEauthorblockA{
\textit{Delft University of Technology}\\
Delft, The Netherlands}
}

\crefformat{chapter}{\S#2#1#3}
\crefmultiformat{chapter}{\S#2#1#3}{ and~\S#2#1#3}{, \S#2#1#3}{, and~\S#2#1#3}

\crefformat{section}{\S#2#1#3}
\crefmultiformat{section}{\S#2#1#3}{ and~\S#2#1#3}{, \S#2#1#3}{, and~\S#2#1#3}

\begin{document}

\maketitle

\begin{abstract}
TLS stripping attacks expose sensitive web traffic by forcing secure HTTPS connections to fall back to unencrypted HTTP.
At present, protection against these attacks relies on website operators explicitly opting into security by deploying mechanisms such as HTTP Strict Transport Security (HSTS) headers.
These mechanisms have significant limitations: some are weak or difficult to configure, which raises the risk of misconfiguration and reduces practical adoption; others violate HTTP backward compatibility; at least one can even be abused to enable unintended user tracking.

We introduce HSTS-Enforced, a mechanism that eliminates the remaining attack surface for TLS stripping while still allowing operators to securely specify that their websites need to be accessed over HTTP when necessary, thereby maintaining accessibility.
To achieve this, we flip the current opt-in security model to an opt-out model: all connections default to HTTPS, and operators can explicitly opt out if their websites require HTTP using so-called HTTP-Required indicators.
We propose two such HTTP-Required indicators: a new DNS record and an HTTP-Required Preload list.
We evaluate HSTS-Enforced under multiple deployment scenarios, demonstrating that it blocks all practical TLS stripping attempts while maintaining compatibility for sites that require HTTP -- without introducing overhead in the typical case. Finally, we outline a practical transition path to accelerate global adoption.
\end{abstract}


\section{Introduction}
The initial milestone in web security was the introduction of \gls{https}~\cite{Rescorla2000}, a protocol that employs \gls{tls} to encrypt communications between clients and servers, ensuring data confidentiality and integrity even if transmissions are intercepted.

Despite this, a practical challenge remains: web clients must implicitly determine whether to initiate a secure \gls{https} connection or fall back to unsecured \gls{http}. A man-in-the-middle attacker can perform \gls{tls} stripping (\Cref{sec:background:attack}), which is an attack that forces connections to use unencrypted \gls{http}.
Several mechanisms exist to protect against \gls{tls} stripping. However, each suffers from notable limitations.
These limitations motivate the design of a more robust and deployable protocol with the following requirements: strong security guarantees, resilience to misconfiguration, ease of achieving secure configurations, interoperability with \gls{http} when needed, and preservation of user privacy (\Cref{sec:background:goals}).
We review existing approaches and analyze the extent to which current protocols fail to satisfy these objectives (\Cref{sec:background:comparison}).

To address the shortcomings of existing approaches, we propose \acrshort{hsts}-Enforced (\Cref{sec:sol:overview}), a mechanism that builds on \gls{hsts}~\cite{Hodges2012}---a policy that instructs web clients to interact with a given website exclusively over \gls{https} for a specified duration---by making this behavior the default for all websites. Rather than requiring operators to opt into this state, web clients default to only creating \gls{https} connections unless website operators explicitly allow the use of \gls{http} by specifying an \gls{http}-Required indicator.
We propose two such \gls{http}-Required indicators (\Cref{sec:sol:indicators}):
(1) a new \gls{dns} record, HTTPREQ, which leverages the cryptographic chain of \gls{dnssec}
to prove that a website requires \gls{http}, and (2) an \gls{http}-Required Preload list, which functions similarly to
\gls{hsts} Preloading but ``in reverse'', i.e., it pre-configures websites to allow \gls{http}.
We also disable \gls{hsts} by default in some special cases where it is convention to do so and propose some additional accessibility features to mitigate the downsides of the secure-by-default behavior of \gls{hsts}-Enforced (\Cref{sec:sol:accessibility}). To further improve the accessibility of websites, we use the additional security provided by \gls{hsts}-Enforced to optimize the connection process of web clients (\Cref{sec:sol:connections}).
We implement \gls{hsts}-Enforced, the proposed indicators, and the optimized connection process in the Chromium browser and several \gls{dns} software suites (\Cref{sec:impl}) and use it to show that \gls{hsts}-Enforced prevents \gls{tls} stripping without inducing significant overhead (\Cref{sec:evaluation}).
Finally, we outline a practical transition path to accelerate global adoption of \gls{hsts}-Enforced (\Cref{sec:transition}).

\section{Background \& Related Work}
\label{sec:background}

\subsection{TLS Stripping Attacks}
\label{sec:background:attack}

\gls{tls} stripping is a downgrade attack in which an adversary forces a web client to communicate over unencrypted \gls{http} even though the target website supports \gls{https}~\cite{Marlinspike2009,Marlinspike2009a}.
The attack exploits the common web client behavior of reverting to 
\gls{http} when an \gls{https} connection attempt fails.
By actively blocking \gls{https} traffic, a man-in-the-middle can prevent the establishment of a \gls{tls} session and coerce the client into using plaintext \gls{http}.

Once the connection is downgraded, the attacker can eavesdrop on sensitive data such as authentication credentials and personal information
or modify the traffic in transit to inject malicious content.
Depending on the attacker’s network position,
these attacks can target all users accessing a specific website or all websites accessed by a particular user or group of users.

\quad\textbf{Threat Model.}\quad
\label{sec:background:threat-model}
We assume an attacker positioned anywhere along the network path between client and server (e.g. a malicious Wi-Fi hotspot operator, a compromised ISP router, or a nation-state interceptor).
The attacker can intercept, block, delay, and replay encrypted traffic; if traffic is unencrypted, they can also read, modify, and forge messages.
The attacker has no direct control over either endpoint: they cannot compromise the user's device or the remote server, install software, or steal cryptographic keys.

This threat model gives the attacker several concrete capabilities.
They can perform \gls{tls} stripping to silently downgrade \gls{https} connections to \gls{http} as described in \Cref{sec:background:attack}.
They can also tamper with \gls{dns} responses, since these travel in plaintext. 
\Gls{dnssec} mitigates this by cryptographically signing responses, making silent modification detectable. However, the attacker can still drop responses to cause lookup failures. \gls{doh} encrypts \gls{dns} queries in transit but does not solve the underlying problem: it merely shifts trust to the resolver, which could return fraudulent responses to the client. Furthermore, \gls{doh} requires the user to both explicitly enable it and select a resolver they trust. We consider this an unreasonable burden to place on general users, who should not need to reason about the trustworthiness of \gls{dns} infrastructure to browse the web securely.

\subsection{The Goal}
\label{sec:background:goals}

We seek a solution that \emph{fully} secures web connections
against \gls{tls} stripping attacks and all subsequent attacks
against \gls{http} by the attacker outlined in our threat model.
The strawman solution would be to never use \gls{http} at all.
However, there are valid reasons for using \gls{http}:
old systems may have performance limitations that impede cryptographic
operations;
website operators may require \gls{http} for testing purposes;
some website operators are simply not willing to acquire an
X.509 certificate;
and finally, local routing devices, such as routers and wireless extenders,
commonly provide a configuration interface that is hosted directly on the
device, and including an X.509 certificate would allow malicious users to
extract the private key linked to the certificate.

\noindent\textbf{Desiderata.}\quad
We derive a set of desiderata
for a mechanism that prevents \gls{tls} stripping attacks:

\noindent\circled[lightgrey]{1} \emph{Maximum security.}
The mechanism should comprehensively prevent \gls{tls} stripping attacks.

\noindent\circled[lightgrey]{2} \emph{Resilience against misconfigurations.}
Incorrect configurations should not lead to a loss of security.

\noindent\circled[lightgrey]{3} \emph{Minimum effort.}
Operators should be able to reach the maximum level of security as easily as possible.

\noindent\circled[lightgrey]{4} \emph{No strict enforcement.}
The mechanism shall allow connections over \gls{http} when explicitly enabled by operators and for some special exceptions, provided this is done in a secure and controlled manner.

\noindent\circled[lightgrey]{5} \emph{No user tracking.}
The mechanism should not enable websites to track users via user-specific responses.

\begin{table}
\caption{Comparison of TLS Stripping Prevention Mechanisms Against Defined Desiderata (\Cref{sec:background:goals}).
}
\label{tab:comparison}
\resizebox{\linewidth}{!}{
\centering
\begin{tabular}{lccccc}
\toprule 
 & \multicolumn{5}{c@{}}{\textbf{Desiderata}} \\
\cmidrule(l){2-6}
\multirow{-2}{*}{\textbf{Mechanism}}
 & \circled[lightgrey]{1}
 & \circled[lightgrey]{2}
 & \circled[lightgrey]{3}
 & \circled[lightgrey]{4}
 & \circled[lightgrey]{5} \\
\midrule 
\gls{hsts} Headers~\cite{Hodges2012}             & \no  & \no  & \no  & \yes & \no  \\
\gls{hsts} Preloading~\cite{ChromiumAuthors2012} & \yes & \no  & \no  & \yes & \yes \\
\gls{https} Records~\cite{Schwartz2023}          & \yes & \no  & \no  & \yes & \yes \\
\gls{https}-First Mode~\cite{HTTPSFirst}         & \no  & \yes & \yes & \yes & \yes \\ 
\gls{https}-Only Mode~\cite{HTTPSOnly}           & \yes & \yes & \yes & \no  & \yes \\ 
\rowcolor[HTML]{f2f2f2} \gls{hsts}-Enforced      & \yes & \yes & \yes & \yes & \yes \\
\bottomrule
\end{tabular}
}
\end{table}

\subsection{Related Work}
\label{sec:background:comparison}

This section mostly focuses on prior solutions that aim to prevent \gls{tls} stripping
and evaluates whether they fulfill our desiderata (\Cref{sec:background:goals}).
\Cref{tab:comparison} summarizes the results.
The remainder of this section addresses work that, while relevant, has a less direct connection to \gls{tls} stripping.

\noindent\textbf{\Gls{hsts} Headers}~\cite{Hodges2012} are communicated as part of an \gls{https} response, after which a web client stores them in their cache.
This enables \gls{hsts} for that website, which enforces connections to a website to use \gls{https} with a trusted certificate.
Nevertheless, \gls{hsts} headers \textit{fail to provide maximum security}\,\circledintext[lightgrey]{1}; although they secure subsequent connections, the first connection to a
website is still vulnerable to attacks.
To prevent accessibility issues, \gls{hsts} headers specify how long they should be cached; after they expire, connections become vulnerable again.
Additionally, the cache is considered part of the client's history; deleting the history clears the stored \gls{hsts} headers.
They are also \textit{not resilient against misconfigurations}\,\circledintext[lightgrey]{2}:
many websites send erroneous \gls{hsts} headers,
causing user agents to not enable \gls{hsts} for these
websites~\cite{Amann2017,Petrov2017,Roth2022,Siewert2022}.
Additionally, configuring \gls{hsts} headers requires more than \textit{minimum effort}\,\circledintext[lightgrey]{3}, causing many websites to not adopt them~\cite{Amann2017,Buchanan2017,Lavrenovs2018,Mendoza2018,Petrov2017,Roth2022,Santos2018}.
Measurement results range from an adoption rate of
0.6\% on a list of websites obtained by scanning IP
addresses for open \gls{https} ports~\cite{Petrov2017}
to 46\% on a list of the top one thousand websites~\cite{Roth2022}.
Finally, because \gls{hsts} headers are part of \gls{https} responses, which can differ per user,
they \textit{unwittingly facilitate user tracking}\,\circledintext[lightgrey]{5}~\cite{Syverson2018}.

\noindent\textbf{\gls{hsts} Preloading}~\cite{ChromiumAuthors2012} relies on the \gls{hsts} Preload list, which is a list maintained by the Chromium project~\cite{ChromiumProjects2024}.
Any website included in the list is pre-configured to always enable \gls{hsts}.
The list removes the need for \gls{hsts} headers to be communicated using \gls{https}, which mitigates some vulnerabilities of \gls{hsts} headers.
However, the list does \textit{not provide resilience against misconfigurations}\,\circledintext[lightgrey]{2} and \textit{fails to require minimum effort}\,\circledintext[lightgrey]{3}:
website operators can request their domain to be added to the \gls{hsts} Preload list using
a registration website maintained by Chromium~\cite{ChromiumAuthors2012},
but the registration process requires correct configuration of website redirects
and \gls{hsts} headers, causing the \gls{hsts} Preload list to see very low adoption rates by website operators~\cite{Petrov2017, Amann2017}.
Through manual scanning, we found that the websites of most major international banks are not on the
list, although banking websites are precisely the ones that should protect against \gls{tls} stripping attacks.
Additionally, if the configuration required for addition to the list is ever broken,
anyone can request the domain's removal.
Currently, there are only around 165.000 domains on the list
that have not broken their configuration~\cite{MozillaPreloadList}, which is a relatively low number considering the size of the Internet.

One additional note is that even when websites are on the \gls{hsts} Preload list,
they remain somewhat vulnerable. To allow greater accessibility any release of the list is only deemed valid for a certain duration. While it is hard for an active attacker to abuse this property, it is more realistic that a user could fail to update their web client for a given duration. The entries on the list would then be considered invalid, which would circumvent its security entirely until the web client is updated.

\noindent\textbf{\gls{https} Records}~\cite{Schwartz2023} are \gls{dns} records that indicate which \gls{https} versions a domain supports.
Theoretically, a \gls{dnssec}-verified negative response for such records could be
treated as an opt-out of \gls{hsts}.
However, since negative \gls{dnssec} responses are automatically generated when a zone is signed, this mechanism would require a broad adoption of \gls{https} records by all websites that do not wish to opt-out, and the accidental removal of these records would lead to a loss of security.
These issues have been shown to be common~\cite{chung2017,dai2016,Shulman2017}.
Therefore, \gls{https} records do not provide \textit{resilience against misconfigurations}\,\circledintext[lightgrey]{2} and do not require \textit{minimum effort}\,\circledintext[lightgrey]{3}.
Similarly, any other \gls{dns} record that operates under an opt-in security model, such as for example using \emph{\gls{dane} records} would fail the same desiderata.

\noindent\textbf{\gls{https}-First Mode}~\cite{HTTPSFirst} is a client-side policy that attempts to establish every connection over 
\gls{https} before falling back to \gls{http}.
This mechanism fails to provide \textit{maximum security}\,\circledintext[lightgrey]{1}. Its security is opportunistic rather than absolute. An active adversary can exploit this by blocking the \gls{https} connection, causing the client to proceed over \gls{http}.

\noindent\textbf{\gls{https}-Only Mode}~\cite{HTTPSOnly} is another feature mostly integrated into browsers.
Previously known as the browser extension \textit{\gls{https} Everywhere}~\cite{EFF2010}, it blocks all \gls{http} connections using a user-facing warning.
Unlike \gls{https}-First mode, \gls{https}-Only mode is not enabled by default in most browsers and can be enabled through the settings if desired.
This mechanism fulfills all desiderata but breaks accessibility on websites that are restricted to \gls{http}. Thus, it fails to \textit{not strictly enforce \gls{https}}\,\circledintext[lightgrey]{4}.
The only difference to our strawman solution (\Cref{sec:background:goals}) is that users can still ignore warnings to access insecure websites. However, doing so effectively leads to the loss of \textit{maximum security}\,\circledintext[lightgrey]{1}.
The primary limitation of this mechanism is its heavy reliance on users to determine whether it is acceptable to disregard a given warning.

\noindent\textbf{Combinations of mechanisms} can be and are being used.
However, for a combination to provide 
\textit{resilience against misconfigurations}\,\circledintext[lightgrey]{2}, require \textit{minimum effort}\,\circledintext[lightgrey]{3}, and \textit{not strictly enforce \gls{https}}\,\circledintext[lightgrey]{4},
all mechanisms used in the combination must fulfill these three desiderata.
There exists no combination of two mechanisms such that both mechanisms fulfill these three desiderata.
Hence, no combination fulfills all five desiderata.

\noindent\textbf{Complementary research} has been done related to \gls{tls} stripping.
A recent proposal extends \gls{hsts} headers to unify them among a group of users in order
to prevent user tracking~\cite{Davitt2024}.
However, this proposal does not address the other issues of \gls{hsts} headers.
Additionally, it is limited to websites with many distinct visitors.

A plethora of work aims to increase web security
in ways that are orthogonal to \gls{hsts}-Enforced,
for instance, by securing certificates~\cite{Abdou2017,Barnes2019,Basin2014,Evans2015,HallamBaker2019,Kim2013,Laurie2021,Tehrani2024}
or by protecting against attacks such as cross-site-scripting~\cite{Balzarotti2008,Barth2011,West2024,Gundy2009,Ren2023}.
Other work focuses on improving privacy and anonymizing users.
Most famously, onion routing~\cite{Dingledine2004} and garlic routing~\cite{I2PProject2024}
enable users to browse the web privately.
The recently proposed Lightweb~\cite{Dauterman2023} anonymizes communication
while protecting against traffic analysis attacks.
Nevertheless,
without \gls{hsts}, one can bypass certificates both in the conventional web and
in exit nodes in onion and garlic routing,
and Lightweb relies on the security of the conventional web to authenticate users.

Several mechanisms have been developed to enhance security in various contexts, such as \gls{dane}~\cite{Hoffman2012}, \gls{mta-sts}~\cite{Margolis2018}, and \gls{sshfp}~\cite{Griffin2006}. None of these mechanisms, however, prevent \gls{tls} stripping attacks. An important thing to note is that their adoption has remained limited, likely because participation is opt-in.

\section{HSTS-Enforced}
\label{sec:solution}
We propose \gls{hsts}-Enforced, a solution where websites default to \gls{hsts}
and can explicitly opt out of security rather than opting in.
In what follows, we provide an overview of \gls{hsts}-Enforced and explain how it fulfills our goals (\Cref{sec:sol:overview}), propose two indicators used to opt out of security \Cref{sec:sol:indicators}), discuss some special cases where we disable \gls{hsts} by default and propose some additional accessibility features (\Cref{sec:sol:accessibility}). Finally, we use the additional security provided by \gls{hsts}-Enforced to optimize the connection process of \gls{hsts} (\Cref{sec:sol:connections}).

\subsection{Overview}
\label{sec:sol:overview}
By default, \gls{hsts}-Enforced enables \gls{hsts} for all websites.
If the use of \gls{http} is desired, operators can explicitly opt out of \gls{hsts}.
By ensuring opt-outs are unspoofable, we achieve
\emph{maximum security}\,\circledintext[lightgrey]{1}.
By defaulting to \gls{hsts}, we \emph{prevent misconfigurations from stifling security}\,\circledintext[lightgrey]{2}.
Moreover, because operators no longer need to explicitly opt into security,
\gls{hsts}-Enforced requires \emph{minimum effort to secure web connections}\,\circledintext[lightgrey]{3}.
\emph{We do not force all connections to use \gls{https}}\,\circledintext[lightgrey]{4}:
websites can indicate that they opt out of security.
Finally, because \gls{hsts}-Enforced prevents user-specific \gls{hsts} settings,
it \emph{prevents websites from using \gls{hsts} state to track users}\,\circledintext[lightgrey]{5}.

In what follows, we provide an overview of our two proposed \gls{http}-Required indicators and explain how they maintain security and prevent introducing new tracking vectors (\Cref{sec:sol:indicators})
To facilitate usability of web clients, we disable \gls{hsts} in some special cases where it is convention to do so and propose some additional accessibility features (\Cref{sec:sol:accessibility}).
Finally, we detail an optimized connection process that minimizes the instances requiring resolution of an indicator and maximizes website accessibility (\Cref{sec:sol:connections}).

\subsection{HTTP-Required Indicators}
\label{sec:sol:indicators}
To opt out of \gls{hsts}, websites should use \gls{http}-Required indicators.
Such indicators must satisfy two demands: (I) \emph{the indicators must not be spoofable}, and (II) \emph{web servers cannot provide a unique (set of) indicator(s) per user}.
These properties prevent the reintroduction of weaknesses that could enable \gls{tls} stripping or user tracking.
We propose two such indicators: the \gls{http}-Required Preload list and the HTTPREQ \gls{dns} record.
When scanning for \gls{http}-Required indicators, the order of checking indicators should be optimized to reduce network traffic and latency.
For our proposed indicators, \gls{hsts}-Enforced first reads the \gls{http}-Required Preload list and then checks for the presence of an HTTPREQ record, as this order minimizes the amount of \gls{dns} traffic.
\Cref{fig:hsts-enabled} illustrates the process of checking whether \gls{hsts}
is enabled or disabled with our proposed indicators.

\noindent\textbf{\gls{http}-Required Preload list.}\quad
This list functions like the \gls{hsts} Preload list in reverse: it enumerates domains (and optionally their subdomains) that opt out of \gls{hsts}.
The list is frequently redistributed and contains an expiration date; entries are re-verified before each release.
We envision the registration service to distribute the latest version of the list every six weeks, 
which matches the current update interval for the \gls{hsts} Preload list.
Operators request inclusion via a registration website; the service verifies from multiple secure vantage points that \gls{http} responses carry an \texttt{HTTP-Required} header, analogously to Let's Encrypt's domain validation~\cite{Aas2020}, which ensures headers are consistently served worldwide and prevents routing-based attacks (I).
Similar to the \gls{hsts} Preload list, the \gls{http}-Required Preload list is integrated directly into clients, preventing user-specific indicators (II).
If the list expires without a client update, access to HTTP-only sites is blocked---a preferable trade-off to compromised security.

\noindent\textbf{HTTPREQ \gls{dns} record.}\quad
We introduce the \emph{``HTTPREQ''} \gls{dns} record, which leverages the cryptographic chain of
\gls{dnssec} to create a secure indicator.
The presence of this \gls{dns} record, combined with a valid \gls{dnssec} signature, serves as the indicator.
Web clients should verify the full cryptographic chain locally in order to make the indicator unspoofable (I).
Using \gls{dns} to distribute this information satisfies demand I, as \gls{dns} tends to propagate records consistently, ensuring that all receivers access the same information.

We allow users to add additional trust anchors for \gls{dnssec} to their web clients that are only valid for a limited set of domains.
Signing an HTTPREQ record using such a custom trust anchor instead of the
original chain of trust allows the record to be accepted as a valid
\gls{http}-Required indicator by anyone who adds the same trust anchor.
With custom trust anchors, system administrators can securely disable \gls{hsts} for a limited group of systems.
To use the custom trust anchors, website operators do not need to add records to the primary \gls{dns} server;
they can configure the \gls{dns} server specified by the network or their systems to mirror the
primary \gls{dns} server and extend it with the required records.
This facilitates use cases such as \gls{http}-based web services inside a secure \gls{vpn} without exposing their usage to potentially malicious outsiders.

\def\flowchartscale{1}
\begin{figure}
    \centering
    \includegraphics[trim=2pt 0pt 6pt 27pt, clip, width=\flowchartscale\linewidth]{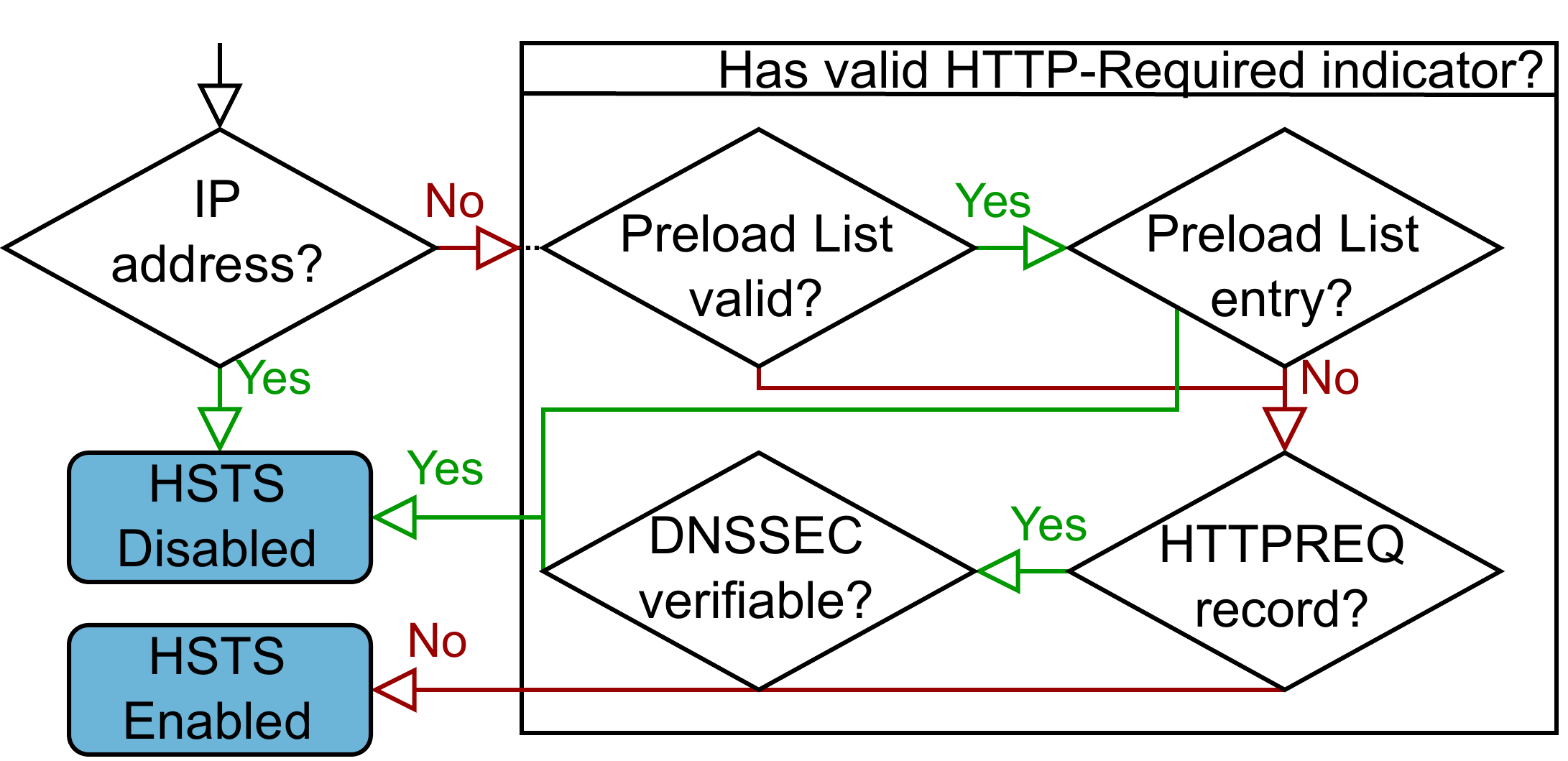}
    \caption{The process of checking whether \gls{hsts} is enabled for a domain with \gls{hsts}-Enforced.}
    \label{fig:hsts-enabled}
\end{figure}

\subsection{Accessibility}
\label{sec:sol:accessibility}
To facilitate accessibility, \gls{hsts}-Enforced disables \gls{hsts} by default in some special cases:
it disables \gls{hsts} for requests that use an IP address (not a domain name), localhost, an \gls{mdns}~\cite{Cheshire2013} domain, or a domain that is reserved for policy or technical
reasons~\cite{IANA2021}, as certificates are not distributed for these addresses.

\noindent\textbf{Local routing devices} (e.g., routers or extenders) require careful consideration.
During initial setup, these devices are often connected directly to a computer or mobile device.
They use this privileged position to override \gls{dns} responses to make their configuration panels accessible via custom domains (e.g., \texttt{fritz.box}).
\gls{hsts}-Enforced connections can block access to panels using this feature, since valid certificates are often missing.
Configuration of an \gls{http}-Required indicator is possible, which would allow these panels to keep functioning normally.
However, some of these domains are not owned by the device vendor, are simultaneously used for public websites, which need to be secure, or exist under unreleased \glspl{tld}.
For domains under unreleased \gls{tld}s, providing an exception is possible, but these exceptions should be actively monitored in case the \gls{tld} is created.
If supported, the configuration panels remain accessible through their \gls{mdns} domain~\cite{Cheshire2013}.
As a last resort, they also remain accessible using their \gls{ip} address. Browsers could assist in identifying the \gls{ip} address of these devices by providing an internal page. By scanning network interfaces and listing default gateways, which share an \gls{ip} address with the configuration panels during initial configuration, a simple overview with links to the panels can be created.

\noindent\textbf{Captive portals} are web pages that users must interact with before getting full access to a public internet connection. These portals should use an \gls{https} login page with a valid certificate and \gls{dns}-based redirects. Alternatively to \gls{dns}-based redirects, OS-level captive portal detection can be used. However, in case a captive portal only supports \gls{http} redirects, users can manually visit a known \gls{http} URL (e.g., \url{http://neverssl.com}) to trigger the portal, as is already required when using a web client that supports \gls{https}-First Mode~\cite{HTTPSFirst}. Although we strongly recommend against using \gls{http} or untrusted \gls{https} for login pages, if providing a trusted \gls{https} connection is not viable, access could be provided via a domain with an \gls{http}-Required indicator, an \gls{mdns} domain, or through the server IP.

\subsection{Connection Process}
\label{sec:sol:connections}
\begin{figure}
    \centering
    \includegraphics[trim=2pt 0pt 2pt 27pt, clip, width=\flowchartscale\linewidth]{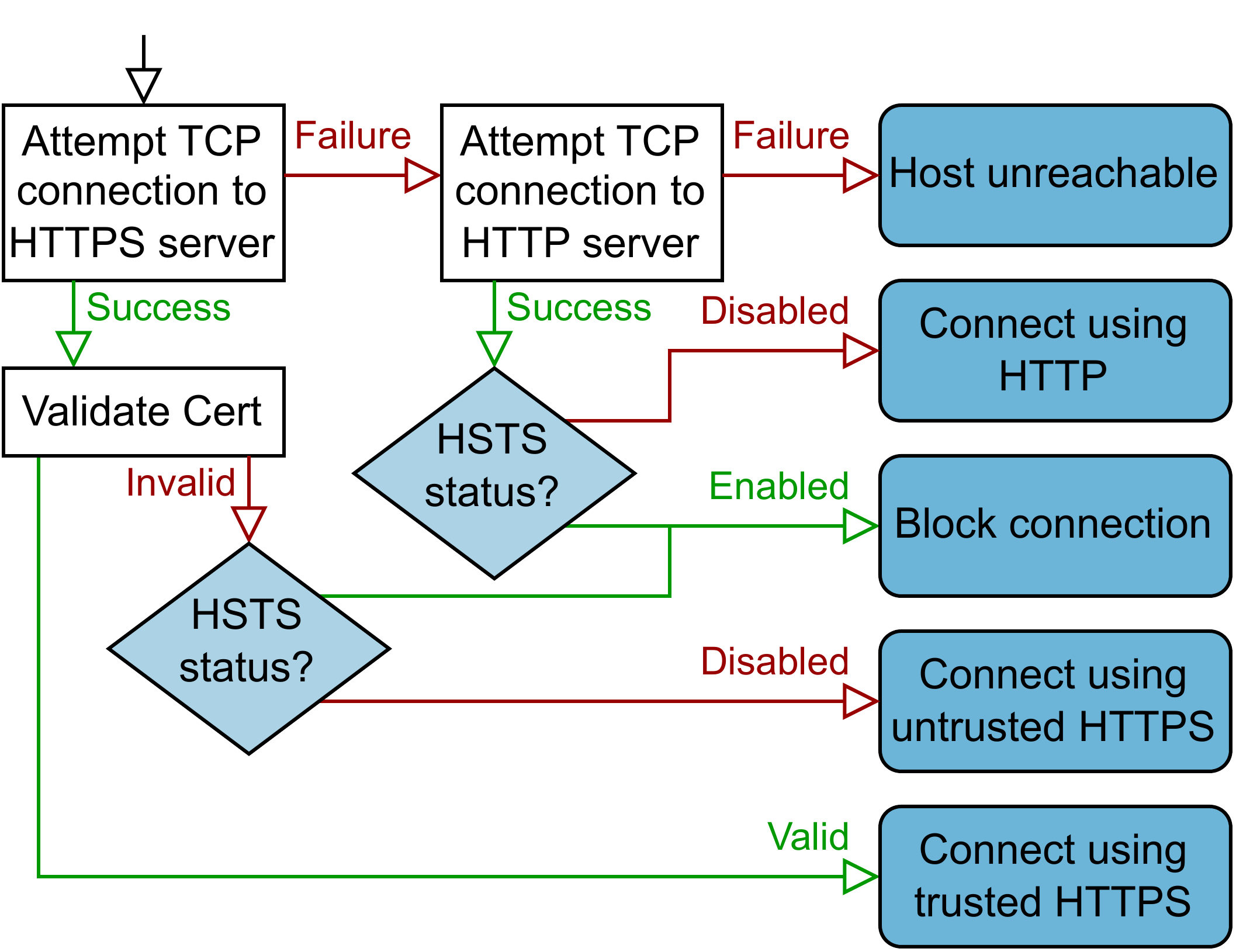}
    \caption{The connection process with \gls{hsts}-Enforced when HTTPS or no protocol are specified in the URL scheme.
    Nodes labeled ``\gls{hsts} status?'' perform the process shown in \Cref{fig:hsts-enabled}.}
    \label{fig:connections-HTTPS}
\end{figure}

\begin{figure}
    \centering
    \includegraphics[trim=2pt 0pt 2pt 27pt, clip, width=\flowchartscale\linewidth]{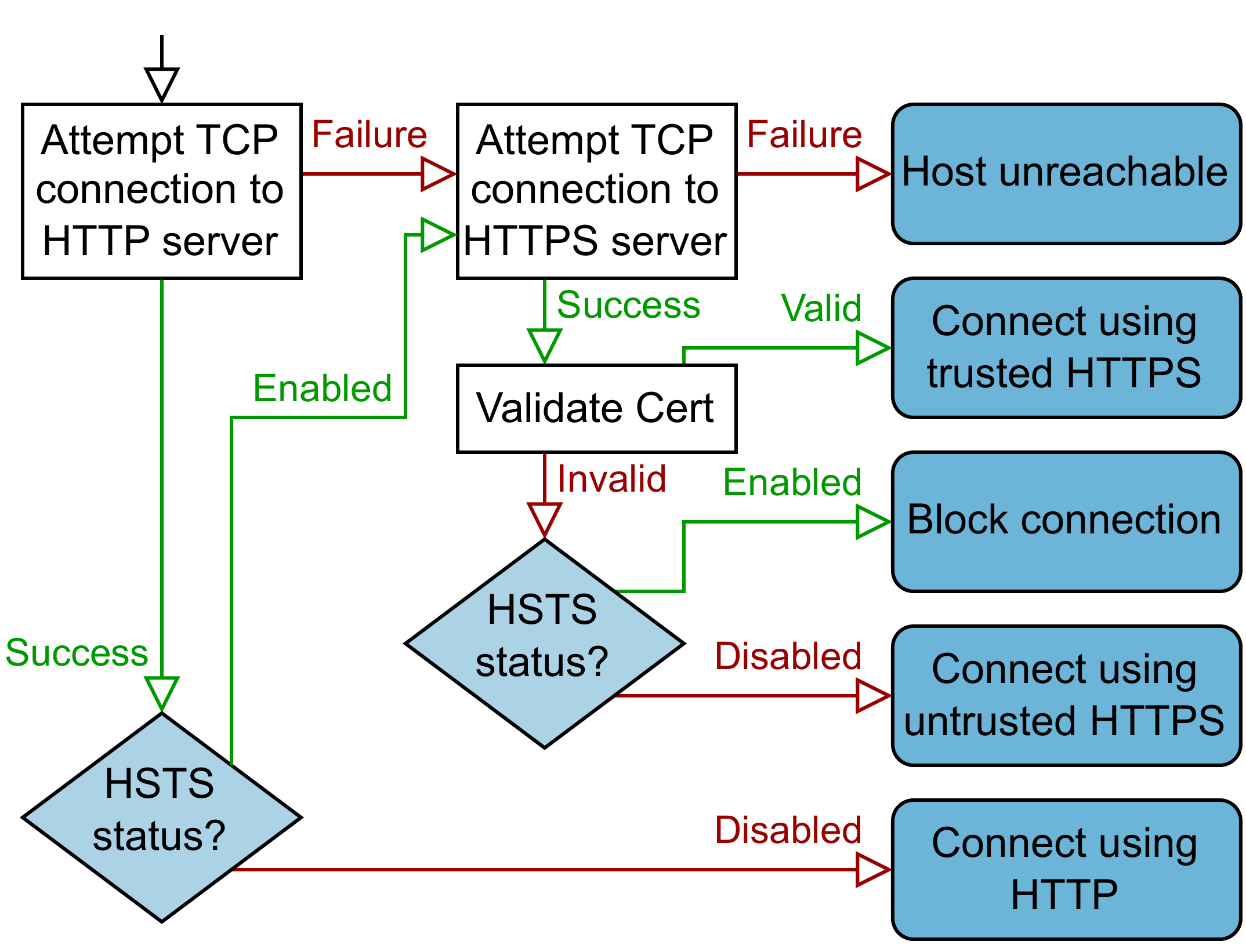}
    \caption{The connection process with \gls{hsts}-Enforced when HTTP is specified in the URL scheme.
    Nodes labeled ``\gls{hsts} status?'' perform the process shown in \Cref{fig:hsts-enabled}.}
    \label{fig:connections-HTTP}
\end{figure}

\gls{hsts}-Enforced decides which connections should be made
based on which scheme is specified in the URL and on the \gls{hsts} state of a website.
In all cases, \gls{hsts}-Enforced delays checking the \gls{hsts} status
until right before making an unsecure connection.
In the common cases of \emph{no scheme} or \emph{\gls{https}}, \gls{hsts}-Enforced always first attempts to establish a connection to the website's \gls{https} server using a trusted certificate and falls back to untrusted \gls{https} and \gls{http} only if \gls{hsts} is disabled (\Cref{fig:connections-HTTPS}).
When \emph{\gls{http}} is explicitly specified in the URL, \gls{hsts}-Enforced follows this directive by first trying to establish an \gls{http} connection if allowed and will always attempt to fall back to \gls{https} if unsuccessful (\Cref{fig:connections-HTTP}).
Attempts to establish a trusted \gls{https} connection will often succeed~\cite{kerschbaumer2025state}.
In case an \gls{https} server is found but the certificate it replies with is untrusted, \gls{hsts}-Enforced will always check whether \gls{hsts} is disabled before establishing such connections.
When attempting an \gls{http} connection, \gls{hsts}-Enforced will first check if an \gls{http} server even exists.
Since \gls{http} relies on TCP, checking if an \gls{http} server exists does not require possibly sensitive data to be sent but only requires an attempt at establishing a connection.
If an \gls{http} server can not be found, \gls{hsts}-Enforced can skip checking the \gls{hsts} status in favor of immediately attempting trusted \gls{https} connections if not attempted before.
If an \gls{http} server is found, it continues by checking the \gls{hsts} status and only allows the connection if it is disabled.
If \gls{hsts}-Enforced cannot reach the \gls{http} server or if \gls{hsts} is enabled, it switches to \gls{https}.

\section{Implementation}
\label{sec:impl}
To demonstrate the feasibility and assess the performance of \gls{hsts}-Enforced, we provide a complete implementation. This implementation includes a modified browser that uses \gls{hsts}-Enforced to initiate web connections and can request and verify the proposed \gls{http}-Required indicators. Additionally, we implement the \gls{dns}-based \gls{http}-Required indicator within a \gls{dns} server and resolver. The full source code is available in the accompanying Git repository~\cite{git}.

\noindent\textbf{Chromium implementation.}\quad
We modified the Chromium browser~\cite{ChromiumProjects2024}
to add support for \gls{hsts}-Enforced.
We removed all code related to \gls{hsts} headers, the \gls{hsts} cache, and the \gls{hsts} Preload list, as it conflicts with \gls{hsts}-Enforced. Afterwards, we added the functionality for checking \gls{http}-Required indicators and modified Chromium's connection setup to follow the procedure described in
\Cref{sec:sol:connections}.

\noindent\textbf{\gls{http}-Required Preload list.}\quad
We implemented a web service for website operators to request addition to the
\gls{http}-Required Preload list, including a REST API and a website with a
graphical user interface.
The service accepts requests for addition and verifies their validity.
It stores accepted entries in a database.
Moreover, a recurring task checks all entries in the database for compliance every six weeks.
Thereafter, it publishes a new version of the list.

We added the \gls{http}-Required Preload list to Chromium.
Its implementation is similar to that of the \gls{hsts} Preload list:
it represents the list as a Huffman-encoded trie~\cite{Huffman1952, Fredkin1960}.
This data structure uses little memory and enables searching the
list in linear time relative to the length of the domain name.

\noindent\textbf{HTTPREQ \gls{dns} record.}\quad
We added support for the HTTPREQ record to three \gls{dns} projects: 
PowerDNS~\cite{PowerDNSTeam2024}, BIND~\cite{ISC2024}, and Unbound~\cite{NLnetLabs2024}. 

We chose PowerDNS for its flexible backend database support, which allows seamless integration with most web-based DNS management panels.
Our BIND implementation serves as a secondary server and provides the \texttt{dig} and \texttt{delv} tools for testing server configurations.
Unbound offers \emph{libunbound}, a library that can be easily integrated into other projects to facilitate \gls{dns} resolution and signature validation. 

To implement HTTPREQ record resolution in Chromium, we integrate the \emph{libunbound} library, as Chromium does not natively support \gls{dnssec} verification and \gls{doh} does not suffice.
We create a single shared \emph{libunbound} context in Chromium's
network process, configured with the default and any user-specified \gls{dnssec} trust anchors.
When the presence of a HTTPREQ record needs to be checked, the connection instantiator uses this \emph{libunbound} context to resolve the record and validate its signature.

\section{Evaluation}
\label{sec:evaluation}
\begin{table*}
\resizebox{\textwidth}{!}{\begin{tabular}{l|c|c|c|c|c|c|c} 
Web server availability                       & \gls{http}                  & untrusted \gls{https}       & \multicolumn{2}{c|}{\gls{http} + untrusted \gls{https}} & \multicolumn{2}{c|}{\gls{http} + \gls{https}} & \gls{https}                 \\ \hline
URL Scheme                                    & \gls{http}/\gls{https}/None & \gls{http}/\gls{https}/None & \gls{http} & \gls{https}/None                           & \gls{http} & \gls{https}/None                & \gls{http}/\gls{https}/None \\ \hline
No valid \gls{http}-Required indicator        & \shield                     & \shield                     & \shield    & \shield                                    & \secure    & \secure                         & \secure                     \\ \hline
$\exists$ valid \gls{http}-Required indicator & \http                       & \untrusted                  & \http      & \untrusted                                 & \http      & \secure                         & \secure
\end{tabular}}
\caption[Results of connection attempts to websites%
    hosting different combinations of web servers,%
    depending on the protocol specified in the URL and%
    whether there is a valid \gls{http}-Required indicator or not. To keep the table compact, columns for which the URL scheme does not affect the outcome have been combined.%
    secure connection indicates a connection using trusted \gls{https}.%
    blocked connection indicates a blocked connection to an unsecure server.%
    http connection indicates a connection using \gls{http}.%
    untrusted connection indicates a connection using untrusted \gls{https}]{%
    Results of connection attempts to websites%
    hosting different combinations of web servers,%
    depending on the protocol specified in the URL and%
    whether there is a valid \gls{http}-Required indicator or not. To keep the table compact, columns for which the URL scheme does not affect the outcome have been combined.%
    \secure~indicates a connection using trusted \gls{https}.%
    \shield~indicates a blocked connection to an unsecure server.%
    \http~indicates a connection using \gls{http}.%
    \untrusted~indicates a connection using untrusted \gls{https}.%
}
\label{tab:effectiveness}
\end{table*}

We verified that \gls{hsts}-Enforced improves web security as intended and typically induces no overhead.
However, upon attempting to visit an unsecure website,
the \gls{dnssec} resolution and validation process can slightly prolong the connection setup.
We detail our experiments hereafter.

\subsection{Setup}
We deployed web servers that support \gls{http}, trusted \gls{https}, or untrusted \gls{https},
as well as recursive and authoritative \gls{dns} servers by deploying them in Docker containers.
Our modified Chromium served as a user agent that ran directly on our system and connected to the servers inside the Docker containers.
We simulated an \gls{rtt} of 20\,ms between the user agent and the recursive \gls{dns} server functioning as their external resolver by adding 10\,ms delay in each direction using netem.
Moreover, we registered a public domain, created and registered the required \gls{dnssec} keys, requested a trusted \gls{https} certificate, and self-signed another certificate to be able to test all aspects of \gls{hsts}-Enforced.

\subsection{Security and Effectiveness}
We verified that \emph{\gls{hsts}-Enforced} \emph{correctly prevents \gls{tls} stripping attacks} by blocking connections that use \gls{http} or untrusted \gls{https} when no valid \gls{http}-Required indicator is present.
It allows connections to such servers if and only if an indicator is present and always allows connections to trusted \gls{https} servers.
The mechanism also respects user preferences: if permitted through an indicator, it employs \gls{http} when explicitly specified through the scheme.
Furthermore, when an indicator is present and the website is only served over a single connection type,
a web client using \gls{hsts}-Enforced consistently uses that connection type, irrespective of the specified scheme.
\Cref{tab:effectiveness} summarizes these results.

Additionally, we verified that adjusting the system time does not compromise the security of \gls{hsts}-Enforced. 
Advancing the system time causes an earlier expiration of the \gls{http}-Required Preload list, which then no longer disables \gls{hsts}. 
Thus, \gls{hsts}-Enforced prevents time-based attacks aimed at breaking \gls{tls} stripping protection.
Without \gls{hsts}-Enforced, such attacks remain feasible against the HSTS Preload list~\cite{Martin2010}.
Failing to update the web client for an extended period results in a security posture comparable to a time-based attack; however, as our evaluation shows, resistance to \gls{tls} stripping remains unaffected in both scenarios.

\subsection{Network Load}
\label{sec:eval:load}
\begin{figure}
    \includegraphics[width=\linewidth]{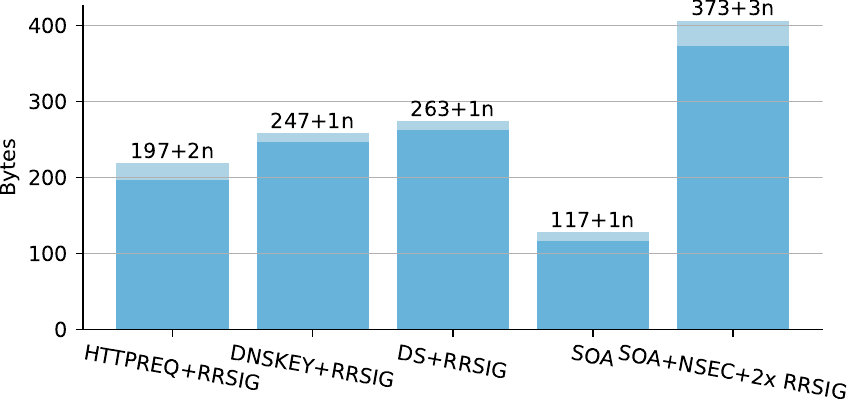}
    \caption{Packet sizes of \gls{dns} records used during resolution of a
    \gls{dnssec}-based \gls{http}-Required Indicator, where $n$ denotes the length of the requested domain name.}
    \label{fig:packet-sizes}
\end{figure}
\gls{hsts}-Enforced causes additional network traffic when verifying an HTTPREQ record.
However,
\emph{its impact is minuscule} because it induces but a few \gls{dns} packets,
which are small in size.
Additionally,
it only induces \gls{dns} traffic in the first place when the website provides \gls{http},
the user explicitly specifies \gls{http},
and no other indicator is specified.
\Cref{fig:packet-sizes} shows the packet sizes of \gls{hsts}-Enforced \gls{dns}
packets.
They are at most around 400 bytes (for a signed negative response with a signed
SOA and NSEC record), but most packets are even smaller.
The size of a signed HTTPREQ response is around 220 bytes.
Signed DNSKEY and DS responses used to verify the signature are around 260 and 270 bytes.

\subsection{Connection Delay}
\begin{figure}
    \includegraphics[width=\linewidth]{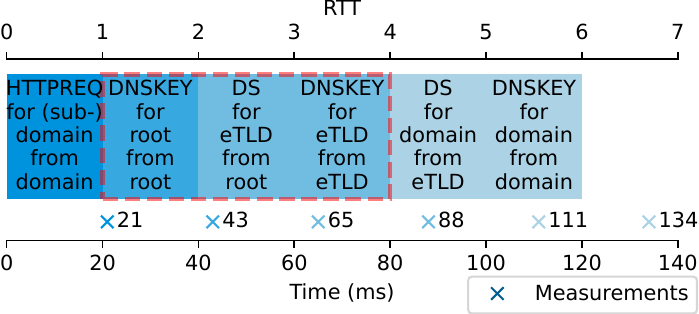}
    \caption{Resolution times for verifying the HTTPREQ record,
    depending on which records are not cached and must be requested.
    Equal colors denote linked records, which are typically both cached if one of them is. The records outlined by the dashed red line are typically already cached due to previous resolutions.}
    \label{fig:resolution-time}
\end{figure}
In most cases, \emph{\gls{hsts}-Enforced does not induce additional delay}.
It only adds some delay when verifying an HTTPREQ indicator
because the \gls{dnssec} resolution can take one or more \glspl{rtt}.
\emph{This delay only occurs when one attempts to initiate an unsecure connection to a website that supports such connections and the domain has an HTTPREQ record}.
It is the result of a trade-off between fast access to an unsecure website and protecting against \gls{tls} stripping attacks.
In the case of a negative response for an HTTPREQ record or any \gls{dnssec} keys, we assume no indicator to be present.
\Cref{fig:resolution-time} illustrates how many \glspl{rtt} the resolution takes
depending on which records must be retrieved and the actual times it took
in our measurements, which are slightly longer than multiples of the
\gls{rtt} due to processing time.
In the worst case, the resolution could take six \gls{rtt}s -- in our measurements,
the worst case took 134\,ms
with an \gls{rtt} of 20\,ms.
This worst case should rarely occur in actuality,
because most of the records that are part of the \gls{dnssec} chain
are cached locally from previous requests.
The likelihood that a record is cached locally increases for records higher up in the chain.
Furthermore, DS and DNSKEY pairs are always used in tandem.
If one of them is cached, usually both are.
In a typical scenario, the DNSKEY for root is already cached
and so are the DS and DNSKEY for the \gls{tld}.
Hence, the typical verification process when visiting an unsecure website the first time should take around three \glspl{rtt} to the resolver.
Additionally, because \gls{dns} response packets are small (a few hundred bytes at most, \Cref{sec:eval:load}), caching many of them is inexpensive.

\subsection{Disk and Memory Usage.}
To ensure compatibility with existing systems, we measured the differences in disk and memory usage of Chromium with and without \gls{hsts}-Enforced.
Both the Debian package size and the memory usage after a clean install vary by less than one percent between the two versions.
We attribute this minimal difference to the fact that the removal of \gls{hsts} headers effectively offsets the inclusion of \emph{libunbound} in both.
Further improvements could be achieved either through deeper integration of \emph{libunbound} or by extending the existing \gls{dns} resolver with \gls{dnssec} support,
which would eliminate the need for \emph{libunbound} entirely.

\section{Transition Strategy}
\label{sec:transition}
We propose a phased transition to \gls{hsts}-Enforced that allows website operators time to adjust before strict enforcement.
During this period, operators of websites that support only \gls{http} connections must deploy one of the \gls{http}-Required indicators or upgrade their site to \gls{https}.
The transition could be supported by a service enabling operators to verify required actions for their sites. Additionally, a user-assisted service could, through browser extensions, gather a list of websites needing configuration updates, allowing a central entity to reach out to their operators. 

After a transitional period, which should be long enough for operators to implement the necessary measures, \gls{hsts}-Enforced can be rolled out via browser updates.
We anticipate its adoption extending beyond web browsers to all web clients that support \gls{http}.
This includes command-line tools (e.g., wget and curl) and web request libraries (e.g., Python’s requests, Java’s HttpClient, Node.js axios, and Go’s net/http), all of which are susceptible to \gls{tls} stripping when an incorrect scheme is specified, because prevailing implementations prioritize the user’s explicit selection of \gls{http} over secure defaults.

\section{Discussion}
\label{sec:discussion}
\noindent\textbf{Adoption of \gls{dnssec}} is not a significant limitation for \gls{hsts}-Enforced.
According to a recent measurement\cite{kerschbaumer2025state}, approximately 92.1\% of web visits are served over \gls{https}, making \gls{hsts} enforcement safe and advantageous for most sites.
\gls{dnssec} only has to be enabled for those websites that can not adopt \gls{https} and want to deploy the \gls{dns} based indicator.
For most of these websites, \gls{dnssec} is a viable option to deploy: all 2820 ICANN-accredited registrars~\cite{ICANN2024a}
and 93.7\% of the \glspl{tld} present in the root zone support \gls{dnssec}~\cite{IANA2024}.
In total, there are 91 \glspl{tld} that do not support \gls{dnssec}.
These \glspl{tld} mostly host only government or organizational sites that likely support \gls{https}.
If deployment of \gls{https} and this indicator is unfeasible, operators can fall back to another indicator such as the HTTP-Required Preload list.

\noindent\textbf{The length of the \gls{http}-Required Preload list} is anticipated to be significantly smaller than the \gls{hsts} Preload list.
The \gls{http}-Required Preload list would primarily cover the limited set of non-local, non-\gls{dnssec} domains that have a hard requirement for \gls{http}.
It is practically impossible to predict the actual length of the list. However, given that these are relatively unpopular \glspl{tld} and only 2.4\% of web visits rely on \gls{http}~\cite{kerschbaumer2025state}, we anticipate that the \gls{http}-Required Preload list may be significantly smaller than the \gls{hsts} Preload list.

\noindent\textbf{\gls{hsts}-Enforced improves upon \gls{https}-First} in two ways. (I) It only falls back to \gls{http} if  a secure indicator is configured. (II) If allowed by the website operator and explicitly indicated by the URL scheme, an \gls{http} connection will be attempted first.
These two changes improve both security and usability of web clients.

\noindent\textbf{\gls{hsts}-Enforced improves upon \gls{https}-Only} by allowing \gls{http} when specified by the website operator through secure indicators without showing an interrupting error page and relying on end-users to make decisions regarding their security, increasing accessibility for the average user without compromising security.

\noindent\textbf{Network-level security mechanisms} remain functional when correctly configured. Intermediary systems, such as \gls{tls} interception, function normally when client devices are configured to trust the necessary certificates. Emergency certificate revocation works reliably to block revoked certificates and replacement certificates are effective as long as the new certificates are valid.

\section{Future work}
\noindent\textbf{Governance of the \gls{http}-Required Preload list} could be decentralized or blockchain-based which would add transparency but also introduce operational complexity. Therefore, governance by a neutral instance such as a standards organization or non-profit Internet governance body with secondary verification by independent entities such as browser developers is desirable.

\noindent\textbf{Alternative \gls{http}-Required indicators} could be added; however, these must fulfill the requirements outlined in \Cref{sec:sol:indicators}.
During the development of \gls{hsts}-Enforced, we evaluated several additional candidate indicators,
such as signed \gls{http} headers, opt-in \gls{https} headers, or an X.509 certificate extension.
The first two of these indicators fail because a web server can vary them per user, enabling potential tracking. X.509 certificate extensions fulfill the requirements; however, it requires an \gls{https} server to serve the X.509 certificate, which is often missing when \gls{http} is needed, making deployment impractical.

\section{Conclusion}
We have described \gls{hsts}-Enforced, a mechanism where web connections
are secured by default, and operators can use simple yet robust methods to opt out
of security if they require \gls{http}.
We proposed two \gls{http}-Required indicators to enable this opt-out while leaving room for additional indicators.
As shown, \gls{hsts}-Enforced secures web connections against attacks that remain possible
with the current system around \gls{hsts} with limited overhead.
We hope that \gls{hsts}-Enforced will be adopted as the next step in securing the web.

\section*{Acknowledgements}
We thank the anonymous reviewers and Chenxing Ji for their comments,
and we acknowledge the open-source tools and frameworks used in this work, whose continued development and maintenance supported the implementation and evaluation of the proposed mechanism.
This research was supported by the National Growth Fund through the Dutch 6G flagship project ``Future Network Services''
and the Netherlands Organisation for Scientific Research (NWO) under the CATRIN project.


\end{document}